\documentclass[aps,prl,reprint]{revtex4-2}
\usepackage{graphicx}
\usepackage{amsmath,amssymb}
\usepackage{bm}
\usepackage[T1]{fontenc}
\usepackage{lmodern}
\usepackage{hyperref}
\usepackage{xcolor}
\usepackage{empheq}
\hypersetup{colorlinks=true,citecolor={blue},linkcolor={blue},urlcolor={blue}}
\usepackage{soul}

\begin{document}

\title{Electrically Tunable and Enhanced Nonlinearity of Moir\'e Exciton-Polaritons in Transition Metal Dichalcogenide Bilayers}

\author{Kok Wee Song}
\affiliation{Department of Physics and Astronomy, University of Exeter, Stocker Road, Exeter EX4 4QL, United Kingdom}
\affiliation{Department of Physics, Xiamen University Malaysia, Sepang 43900, Malaysia}

\author{Oleksandr Kyriienko}
\affiliation{Department of Physics and Astronomy, University of Exeter, Stocker Road, Exeter EX4 4QL, United Kingdom}
\affiliation{School of Mathematical and Physical Sciences, University of Sheffield, Sheffield S10 2TN, United Kingdom}

\begin{abstract}
    We develop a microscopic theory for nonlinear optical response of moir\'e exciton-polaritons in bilayers of transition metal dichalcogenides (TMDs). Our theory allows to study the tunnel-coupled intralayer and interlayer excitonic modes for a wide range of twist angles ($\theta$), external electric field, and light-matter coupling, providing insights into the hybridization regime inaccessible before. Specifically, we account for the Umklapp scattering processes of two exciton-polaritons responsible for enhanced nonlinearity, and show that it is crucial for describing interactions at strong hybridization. We reveal a regime of attractive nonlinearity for moir\'e polaritons, stemming from the anisotropic Coulomb interactions, which can explain some of experimental features of optical response in TMD bilayers. Furthermore, within our theory we demonstrate that the attractive nonlinearity can be tuned into repulsive by applying an external electric field. Our findings show that nonlinear moir\'e polaritons offer a controllable platform nonlinear polaritonic devices.
\end{abstract}

\maketitle

\emph{Introduction.---}Moir\'e superlattices formed in twisted bilayers of atomically thin crystals represent a unique platform for studying strongly correlated physics \cite{Balents2020,Carr:NatRevPhys5-2020,Bloch2022}, with emergent superconductivity in moir\'e bilayers of graphene serving as a prominent example \cite{Cao:Nature556-2018}. The essence of moir\'e engineering is to generate the band mixing between monolayer crystals by forming a moir\'e pattern with reduced crystal translational symmetry. At small twist angles, as the folded band energies come closer together, the interlayer tunneling induces a strong band hybridization that profoundly changes low energy states of materials, manifested in flatbands and unique transport properties \cite{Wu:PRL121-2018,Tang:Nature579-2020,Regan:Nature579-2020}. Moir\'e engineering is also an effective tool for tuning the optical properties \cite{Wu:PRB97-2018,Tran:Nature567-2019,Tran:2DMat8-2020,Campbell:NatPhys20-2024}. This was demonstrated in TMD bilayers, revealing their stacking- and twist-dependent optics \cite{Shinde:NPGAsiaMat10-2018,Li:PRB106-2022,Seyler:Nature567-2019,Zhang:NatCommun11-2020,Paradisanos:PRB105-2022,Villafane:PRL130-2023}.
\begin{figure}
    \centering
    \includegraphics[width=3.35in]{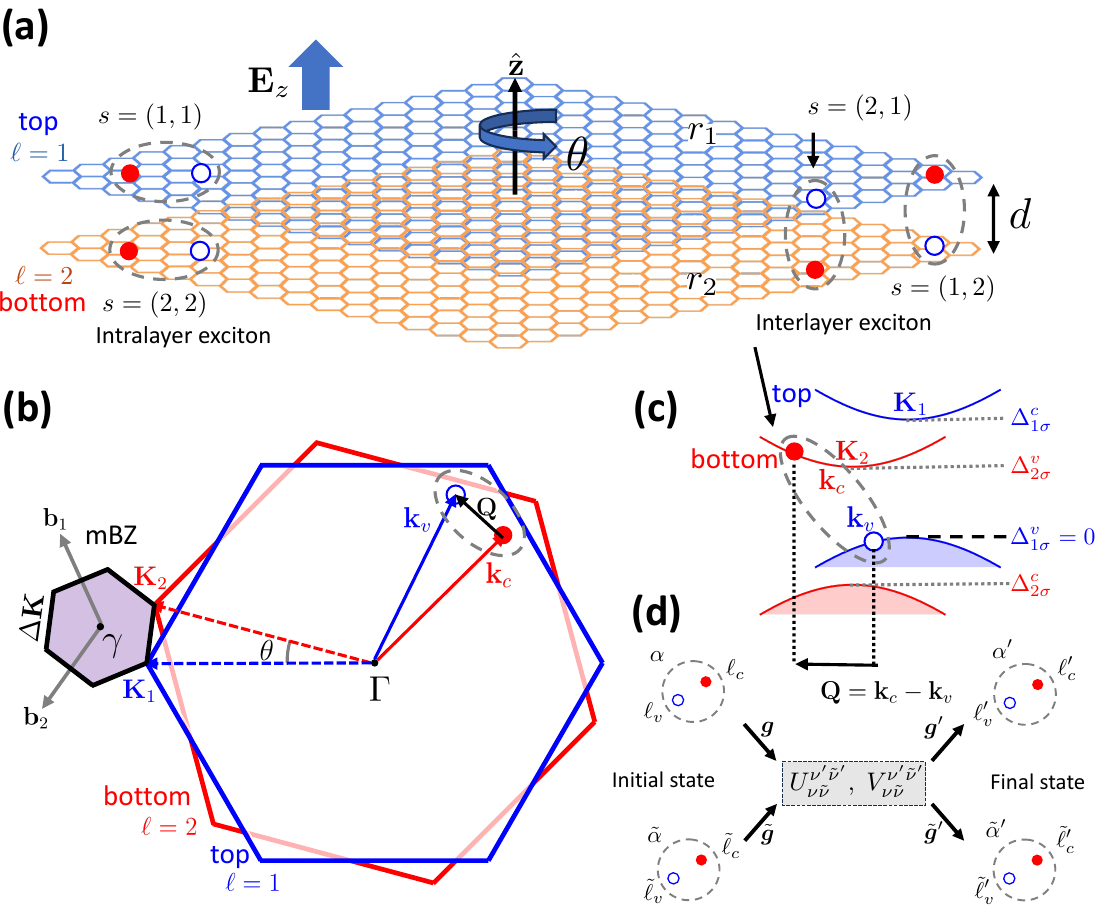}
    \caption{(a) Sketch of a bilayer system, showing two types of interlayer and intralayer excitons. (b) Brillouin zones of top (blue) and bottom (red) monolayer. Mini Brillouin zones (mBZ) of the moir\'e bilayer. (c) Sketch of interlayer exciton in momentum space. (d) Diagram of exciton-exciton Coulomb interaction where electron and hole layer indices $\ell$ do not change, and same-layer identical particles can be exchanged.}
    \label{fig:MoireX}
\end{figure}

Moir\'e excitons, represented by electron-hole bound states spread over several layers of material, lead to a pronounced optical response of 2D bilayers \cite{Tartakovskii:NatMat19-2020,Huang:NatNanotech17-2022}, and offer a platform for quantum optical applications \cite{Keogl:npg2DMat7-2023,Sun:ChemRev124-2024}. Their linear properties demonstrate a dipolar response and electrical tunability \cite{Klein:ACSnanolett17-2017,Yu:SciBull65-2020,Tang:NatNanoTech16-2021,Sponfeldner:PRL129-2022,Kovalchuk:arXiv03-2023}. Similar hybridized intra-interlayer excitons appear in homobilayer systems for selected stacking \cite{Gerber:PRB99-2019,Paradisanos2020,Peimyoo2021,Charalambos:NatCommun14-2023}. When embedded into a microcavity, hybridized moir\'e excitons can couple strongly to photons and form polaritons \cite{Zhang:Nature591-2021,Datta2022,Charalambos:NatCommun14-2023}. Nonlinear optical response of hybridized excitons and polaritons in TMDs revealed enhancement as compared to the monolayer response \cite{Kremser:npj2DMat4-2020,Zhang:Nature591-2021,Datta2022,Charalambos:NatCommun14-2023,Hu:Science380-2023} and suggested electric field dependence \cite{Fitzgerald:NanoLett22-2022,Erkensten:Nanoscale15-2023}. The presence of nonlinearity is important for the emergence of exotic correlated phases \cite{Shimazaki:Nature580-2020,Baek:NatNanotech16-2021,Zimmerman:PNAS119-2022,Campbell:npj2DMatApp6-2022,HuangHafezi2024} and underpins strong photonic nonlinearity \cite{Zhang:Nature591-2021,Song:PRR6-2024,Camacho-Guardian:PRL128-2022} important for driving the system into a quantum regime \cite{Wang2024,Thureja2022,HuChervy2024}. However, despite significant advances, the theoretical description of nonlinear interaction for hybridized moir\'e-type excitons and polaritons has been limited to specific stackings (thus not twist-dependent), considering simplified mode wavefunctions, and treating fixed (top-bottom) excitonic configurations.

In this Letter, we develop a microscopic theory of twisting-dependent optical response of 2D bilayers, taking into account a rich structure of moir\'e modes. Solving a Wannier equation within the multi-Gaussian basis expansion, we resolve excitonic modes emerging from hybridization of interlayer and intralayer states at varying twisting angles and external bias. Crucially, this allows calculating nonlinear scattering for different modes without simplifying their structure, as well as describing strong light-matter coupling. We observe that the Umklapp scattering involving neighboring mini Brillouin zones (mBZ) of the moir\'e lattice play an important role for nonlinearity, leading to its enhancement for non-zero $\theta$. For bilayers with small twisting we reveal an attractive nonlinearity driven by exchange processes, and recover a conventional dipolar repulsion for large ($>3^{\circ}$) twists. Our results can offer microscopic insights into nonlinear redshifts measured experimentally \cite{Charalambos:NatCommun14-2023,Steinhoff:PRX14(2024)}, and open avenues for quantitative studies of quantum moir\'e polaritons.

\emph{Model.---}We consider a bilayer Hamiltonian $\mathcal{H}=\mathcal{H}_0+\mathcal{H}_{\mathrm{int}}+\mathcal{H}_t$, consisting of the free energy terms, Coulomb interaction term, and the tunneling term, respectively. The free energy for electrons (e) and hole (h) in different layers reads as $\mathcal{H}_0=\sum_{\mathbf{k}\sigma}\sum_{\ell=1,2}[\varepsilon^{c}_{\ell\sigma}(\mathbf{k})a^{\ell,\dagger}_{ \mathbf{k}\sigma}a^{\ell}_{\mathbf{k} \sigma}+\varepsilon^{v}_{\ell\sigma}(\mathbf{k})b^{\ell,\dagger}_{ \mathbf{k}\sigma}b^{\ell}_{\mathbf{k}\sigma}]$, where a crystal momentum $\mathbf{k}$ is measured from the $\Gamma$-point in each layer labeled as $\ell=1,2$, and $\sigma=\uparrow,\downarrow$ is the spin index. The density-density interaction term reads $\mathcal{H}_{\mathrm{int}} = \frac{1}{2L^2}\sum_{\ell\ell',\mathbf{q}}v_{\ell\ell'}(\mathbf{q})
\rho_{\ell\mathbf{q}}\rho_{\ell',-\mathbf{q}}$, where $L^2$ is a sample area. Since only the low-energy e-h pairs are relevant for forming the exciton-bound state \cite{Latini:PRB92-2015,Ceferino:PRB101-2020}, we expand the dispersion near the valleys at $\mathbf{K}_\ell$ (band edge) as $\varepsilon^{\mathrm{b}}_{\ell\sigma}(\mathbf{k})=\Delta^{\mathrm{b}}_{\ell\sigma}\pm(\mathbf{k}-\mathbf{K}_{\ell})^2/(2m_{\ell}^\mathrm{b})$ with $\mathrm{b}=c,v$ being the conduction and valence band index, and  $m^{\mathrm{b}}_\ell$ being the $\mathrm{b}$-band mass in layer $\ell$. We let the energy offset for the $\mathrm{b}$-band edge
as $\Delta^{\mathrm{b}}_{\ell\sigma}$ that is measured from the topmost valence band (see Fig.~\ref{fig:MoireX}c). In the Coulomb term, $\rho_{\ell\mathbf{q}}=\sum_{\mathbf{k}\sigma}(a^{\ell\dagger}_{ \mathbf{k}+\mathbf{q},\sigma}a^{\ell}_{\mathbf{k} \sigma}+b^{\ell\dagger}_{ \mathbf{k}+\mathbf{q},\sigma}b^{\ell}_{\mathbf{k} \sigma})$ is the charge density, operator and the screened potential between electrons is of the Keldysh-Rytova form \cite{Danovich:PRB97-2018},
\begin{equation}
v_{\ell\ell'}(\mathbf{q})=\frac{2\pi}{\epsilon q}
\frac{\kappa_{\ell\ell'}(q)}{(1+r_{1} q)(1+r_{2} q)-r_{1}r_{2}q^2\mathrm{e}^{-2qd}} ,
\label{eqn:Keldysh}
\end{equation}
where $\epsilon$ is the dielectric constant of the environment, $r_\ell$ is the screening length of the material, and $d$ is the interlayer distance (see Fig.~\ref{fig:MoireX}a). Here, $\kappa_{12}(q)=\kappa_{21}(q)=\mathrm{e}^{-qd}$, $\kappa_{11}(q)=1+r_2q(1-\mathrm{e}^{-2qd})$, and $\kappa_{22}(q)=1+r_1q(1-\mathrm{e}^{-2qd})$. The interlayer hopping Hamiltonian in the continuum limit reads as
\begin{align}
\mathcal{H}_{t}\!=\!\sum_{\ell\ell'}\sum_{\mathbf{k}\mathbf{k}'}&\Big[T^c_{\ell'\ell}(\mathbf{k}',\mathbf{k})a^{\ell',\dagger}_{\mathbf{k}'\sigma}a^{\ell}_{\mathbf{k}\sigma}\!+\!T^v_{\ell'\ell}(\mathbf{k}',\mathbf{k})b^{\ell',\dagger}_{\mathbf{k}'\sigma}b^{\ell}_{\mathbf{k}\sigma}\Big]\label{eqn:Ht}
\end{align}
with interband hoppings being omitted \cite{Ruiz-Tijerina:PRB99-2019}.  
 
\emph{Exciton bound states.---}The Coulomb interaction described by $\mathcal{H}_{\mathrm{int}}$ gives rise to an exciton bound state described by a linear combination
\begin{equation}\label{eqn:X}
    X^{s\dagger}_{\alpha\sigma}(\mathbf{Q})=\sum_{\mathbf{k}_c,\mathbf{k}_v}\delta_{\mathbf{k}_c-\mathbf{k}_v,\mathbf{Q}}\phi^{s}_{\alpha\sigma}(\mathbf{p})
a^{\ell_c\dagger}_{\mathbf{k}_c\sigma}b^{\ell_v}_{\mathbf{k}_v\sigma},
\end{equation}
where $\alpha$ is the exciton principal quantum number, and excitonic wavefunction is described by separated variables corresponding to the center-of-mass motion with total momentum $\mathbf{Q}=\mathbf{k}_c-\mathbf{k}_v$, and the relative motion with momentum $\mathbf{p}=[m^v_{\ell_v}(\mathbf{k}_c-\mathbf{K}_{\ell_c})+m^c_{\ell_c}(\mathbf{k}_v-\mathbf{K}_{\ell_v})]/(m^c_{\ell_c}+m^v_{\ell_v})$. 
Here, $s=(\ell_c,\ell_v)$ is the double index labeling the exciton species (intralayer/interlayer, see Fig.~\ref{fig:MoireX}a). 
The relative motion wavefunction satisfies the Wannier equation
\begin{align}\label{eqn:X-eqn}
\Big[\frac{\mathbf{p}^2}{2\mu_{s}}+\frac{(\mathbf{Q}-\Delta\mathbf{K})^2}{2M_{s}}+\varepsilon^{s}_\sigma\Big]\phi^{s}_{\alpha\sigma}(\mathbf{p})
-&\sum_{\mathbf{q}}v_{s}(\mathbf{q})\phi^{s}_{\alpha\sigma}(\mathbf{p}+\mathbf{q})
\notag\\
    =&E^{s}_{\alpha\sigma}(\mathbf{Q})\phi^{s}_{\alpha\sigma}(\mathbf{p}),
\end{align}
with $\Delta\mathbf{K}=\mathbf{K}_{\ell_c}-\mathbf{K}_{\ell_v}$, $\varepsilon^s_{\sigma}=\Delta^{c}_{\ell_c\sigma}-\Delta^{v}_{\ell_v\sigma}$, and $E^{s}_{\alpha\sigma}(\mathbf{Q})$ being the exciton energy. The total mass and reduced mass are $M_{s}=m^c_{\ell_c}+m^v_{\ell_v}$, and  $\mu_{s}=m^c_{\ell_c}m^v_{\ell_v}/M_s$. 

\emph{Hybridized moir\'e exciton.---}Due to the interlayer hopping, the exciton in the moir\'e bilayer becomes a hybridized state between different excitonic modes in Eq.~\eqref{eqn:X} with $(s,\alpha,\mathbf{Q})$. To find the \emph{hybridized} moir\'e exciton, we approximate the interlayer hopping constant in the vicinity of the valleys $\mathbf{K}_{\ell}$ in both layers as
\cite{Wang:PRB95-2017,Ruiz-Tijerina:PRB99-2019}
\begin{align}\label{eqn:Tc_Tv}
    T^{c,v}_{\ell'\ell}(\mathbf{k}',\mathbf{k})=&\sum_{\mathbf{G}_{\ell}\mathbf{G}_{\ell'}}t^{c,v}(\mathbf{K}_\ell+\mathbf{G}_{\ell})\delta_{\mathbf{k}'-\mathbf{k},\mathbf{G}_{\ell'}-\mathbf{G}_{\ell}}\tau^x_{\ell\ell'}, 
\end{align}
where $\mathbf{G}_{\ell}$ is the monolayer reciprocal lattice vector in layer $\ell$. The matrix $\tau^x_{11}=\tau^x_{22}=0$ and $\tau^x_{12}=\tau^x_{21}=1$. 
This interlayer scattering reduces the translational crystal symmetry of the monolayer lattice into the moir\'e superlattice. As indicated by the delta function in Eq.~\eqref{eqn:Tc_Tv}, this scattering preserved the momentum up to $\mathbf{G}_{\ell}-\mathbf{G}_{\ell'}=i\mathbf{b}_1+j\mathbf{b}_2$
where $i$, $j$ are integers and $\mathbf{b}_{1,2}$ are the primitive reciprocal lattice vectors of the moir\'e bilayer \cite{Wang:PRB95-2017} (see Fig.~\ref{fig:MoireX}b). As a result, the momentum within the moir\'e mini Brillouin zone $\bar{\mathbf{Q}}$ is a conserved quantity. 
Therefore, this $\bar{\mathbf{Q}}$-preserving interlayer scattering  leads to the formation of a \emph{hybridized} moir\'e exciton as 
\begin{equation}\label{eqn:X-mix}
    \mathcal{X}^{\bar{\alpha}\dagger}_{\sigma}(\bar{\mathbf{Q}})=\sum_{\nu}C^{\bar{\alpha}}_{\nu\sigma}(\bar{\mathbf{Q}})X^{\dagger}_{\nu\sigma}(\bar{\mathbf{Q}}),
\end{equation}
where we have used the shorthand index $\nu=(s,\alpha,\bm{g})$ to represent exciton species $s$, exciton state $\alpha$, and reciprocal lattice vector $\bm{g}=i\mathbf{b}_1+j\mathbf{b}_2$ to lighten our notation (Fig.~\ref{fig:MoireX}d). Also, we let $X^\dagger_{\nu\sigma}(\bar{\mathbf{Q}})=X^{s\dagger}_{\alpha\sigma}(\bar{\mathbf{Q}}+\bm{g})$. The \emph{hybridized} moir\'e exciton can be obtained by solving
\begin{equation}\label{eqn:hX-eqn}
        [E_{\nu\sigma}(\bar{\mathbf{Q}})-\mathcal{E}^{\bar{\alpha}}_{\sigma}(\bar{\mathbf{Q}})]C^{\bar{\alpha}}_{\nu\sigma}(\bar{\mathbf{Q}})
\!=\!
\sum_{\nu'}w_{\nu'\nu}(\bar{\mathbf{Q}})C^{\bar{\alpha}}_{\nu'\sigma}(\bar{\mathbf{Q}})
\end{equation}
where $E_{\nu\sigma}(\bar{\mathbf{Q}})=E^s_{\alpha\sigma}(\bar{\mathbf{Q}}+\bm{g})$, and the interlayer-to-intralayer exciton transition matrix elements is
\begin{equation}
    \langle 0|X_{\nu'}(\bar{\mathbf{Q}}')\mathcal{H}_tX^{\dagger}_{\nu}(\bar{\mathbf{Q}})|0\rangle=w_{\nu'\nu}(\bar{\mathbf{Q}})\delta_{\bar{\mathbf{Q}}',\bar{\mathbf{Q}}}
\end{equation}
with $|0\rangle$ being the ground state. In Eq.~\eqref{eqn:hX-eqn}, we label the \emph{hybridized} moir\'e exciton eigenstates by index $\bar{\alpha}$ with a bar, see details in Supplemental Material (SM) \cite{SM}. The excitonic model here is different from the moir\'{e} potential approach\cite{Wu:PRL121-2018,Wu:PRB97-2018,HuangHafezi2024} which does not account for the interlayer-intralayer exciton hybridization.


\emph{Exciton-exciton interaction.---}In our analysis we go beyond single-particle properties and study correlation effects for moir\'e excitons arising from their interactions. We concentrate on 1$s$ states ($\alpha=0$) and set $\bar{\mathbf{Q}}=0$ in the scattering processes, such that we characterize the low-energy exciton-exciton (X-X) interactions with elastic scattering for $\bar{\mathbf{Q}}=0$ only. Focusing on low-density regime, the X-X interaction between $\mathcal{X}^{\bar{\beta}\dagger}$ and $\mathcal{X}^{\bar{\alpha}\dagger}_{\sigma}$ with states $\bar{\alpha}$ and $\bar{\beta}$ can be calculated from the total energy of the two-exciton state within the same valley $\Omega^{\bar{\beta}\bar{\alpha}}_{\sigma}=\langle0|\mathcal{X}^{\bar{\alpha}}_{\sigma}\mathcal{X}^{\bar{\beta}}_{\sigma}\mathcal{H}\mathcal{X}^{\bar{\beta}\dagger}_{\sigma}\mathcal{X}^{\bar{\alpha}\dagger}_{\sigma}|0\rangle=\mathcal{E}^{\bar{\beta}}_{\sigma}+\mathcal{E}^{\bar{\alpha}}_{\sigma}+\Delta^{\bar{\beta}\bar{\alpha}}_{\sigma}$. 
The interacting potential energy is given by
\begin{align}
\Delta^{\bar{\beta}\bar{\alpha}}_{\sigma}=\sum_{\nu\nu'}\sum_{\tilde{\nu}\tilde{\nu}'}&\bar{C}^{\bar{\beta}}_{\nu'\sigma}\bar{C}^{\bar{\alpha}}_{\tilde{\nu}'\sigma}C^{\bar{\beta}}_{\nu\sigma}C^{\bar{\alpha}}_{\tilde{\nu}\sigma}\delta_{\tilde{\bm{g}}'+\bm{g}',\tilde{\bm{g}}+\bm{g}}\notag\\
&
\Big[U_{\nu\tilde{\nu}}^{\nu'\tilde{\nu}'}+U_{\nu\tilde{\nu}}^{\tilde{\nu}'\nu'}-V_{\nu\tilde{\nu}}^{\nu'\tilde{\nu}'}-V_{\nu\tilde{\nu}}^{\tilde{\nu}'\nu'}\Big]\label{eqn:Deltamn}
\end{align}
where the direct \cite{Erkensten:Nanoscale15-2023} and exchange interactions\cite{Combescot:PhysRep5-2008} are
\begin{align}
U_{\nu\tilde{\nu}}^{\nu'\tilde{\nu}'}=&\sum_{\mathbf{k}\tilde{\mathbf{k}}\mathbf{q}}\Gamma_{\nu\tilde{\nu}}^{\nu'\tilde{\nu}'}(\mathbf{k}\tilde{\mathbf{k}},\mathbf{q})\delta_{\tilde{\ell}_c\tilde{\ell}_c'}\delta_{\tilde{\ell}_v\tilde{\ell}_v'}\delta_{\ell_c\ell_c'}\delta_{\ell_v\ell_v'}\delta_{\mathbf{q},\bm{g}'-\bm{g}},
    \notag\\ V_{\nu\tilde{\nu}}^{\nu'\tilde{\nu}'}=&\sum_{\mathbf{k}\tilde{\mathbf{k}}\mathbf{q}}\Gamma_{\nu\tilde{\nu}}^{\nu'\tilde{\nu}'}(\mathbf{k}\tilde{\mathbf{k}},\mathbf{q})\delta_{\tilde{\ell}_c\ell_c'}\delta_{\tilde{\ell}_v\tilde{\ell}_v'}\delta_{\ell_c\tilde{\ell}_c'}\delta_{\ell_v\ell_v'}\delta_{\mathbf{q},\bm{g}'-\bm{g}+\mathbf{k}-\tilde{\mathbf{k}}},\notag
\end{align}
that are depicted in Fig.~\ref{fig:MoireX}d. Here, $\mathbf{q}$ is the transferred momentum between $\bar{\alpha}$- and $\bar{\beta}$-exciton and the scattering potential is $\Gamma_{\nu\tilde{\nu}}^{\nu'\tilde{\nu}'}(\mathbf{k}\tilde{\mathbf{k}},\mathbf{q})=\sum_{\ell\ell'}f^{\ell}_{\nu\sigma}(\mathbf{k},\mathbf{q}) (v_{\ell\ell'}(\mathbf{q})/2L^2) f^{\ell'}_{\tilde{\nu}\sigma}(\tilde{\mathbf{k}},-\mathbf{q})\phi^{\ast}_{\nu'}(\mathbf{k})\phi^{\ast}_{\tilde{\nu}'}(\tilde{\mathbf{k}})$, 
%
%
with excitonic wavefunction being expressed in $\nu$-index notation as $\phi_{\nu\sigma}(\mathbf{k})=\phi^{s}_{\alpha\sigma}(\mathbf{k}-m^c_{\ell_c}/M_{s}\bm{g})$, and the factor
$f^{\ell}_{\nu\sigma}(\mathbf{k},\mathbf{q})=\delta_{\ell_c\ell}\phi_{\nu\sigma}(\mathbf{k}-\mathbf{q})-\delta_{\ell_v\ell}\phi_{\nu\sigma}(\mathbf{k})$. The calculation of the direct interaction $U_{\nu\tilde{\nu}}^{\nu'\tilde{\nu}'}$ is straightforward while evaluating the exchange interaction  $V_{\nu\tilde{\nu}}^{\nu'\tilde{\nu}'}$ is rather involved and is detailed in SM~\cite{SM}. We remark that, the result in Eq.\eqref{eqn:Deltamn} is for intravalley interaction and only valid for the low-density regime.

\emph{Moir\'e exciton polariton.---}Embedding the moir\'e bilayer into an optical microcavity, excitons can couple strongly to cavity photons, forming polaritonic states. This is an essential mechanism for studying optical nonlinearity \cite{Kuriakose2022,Makhonin:LightSciAppl13(2024)}. Here, dipolar polaritons in GaAs double quantum wells serve as an inspiration for studying pronounced nonlinear effects \cite{Rosenberg2018,Togan2018,Kyriienko2014,Kyriienko2016}. To study moir\'e polaritons, we introduce light-matter coupling as an additional term in the system Hamiltonian corresponding to $\mathcal{H}_{\mathrm{sc}} = \sum_{\ell}d^\ell_{cv}\sum_{\mathbf{k}\mathbf{Q}}c^\dagger_{\mathbf{Q}}a^\dagger_{\ell\mathbf{k}} b_{\ell\mathbf{k}+\mathbf{Q}}+h.c.= \sum_{\bar{\alpha}\bar{\mathbf{Q}}}g^{\bar{\alpha}}_c(\bar{\mathbf{Q}})c^\dagger_{\bar{\mathbf{Q}}}\mathcal{X}^{\bar{\alpha}}(\bar{\mathbf{Q}})+h.c.$
Here $c^\dagger_{\mathbf{Q}}$ is the photonic field operator, and  $d_{cv}^\ell$ is the interband transition matrix element in $\ell$-th layer. In the equation, this coupling term is written in the hybridized excitonic basis with coupling constant $ g^{\bar{\alpha}}_c(\bar{\mathbf{Q}})=\sum_{\ell\alpha\mathbf{k}}d_{cv}^{\ell}\phi^{\ell\ell}_{\alpha}(\mathbf{k})\bar{C}^{\bar{\alpha},\ell\ell}_{\alpha\sigma}(\bar{\mathbf{Q}})$.
\begin{figure}
    \centering
    \includegraphics[width=3.2in]{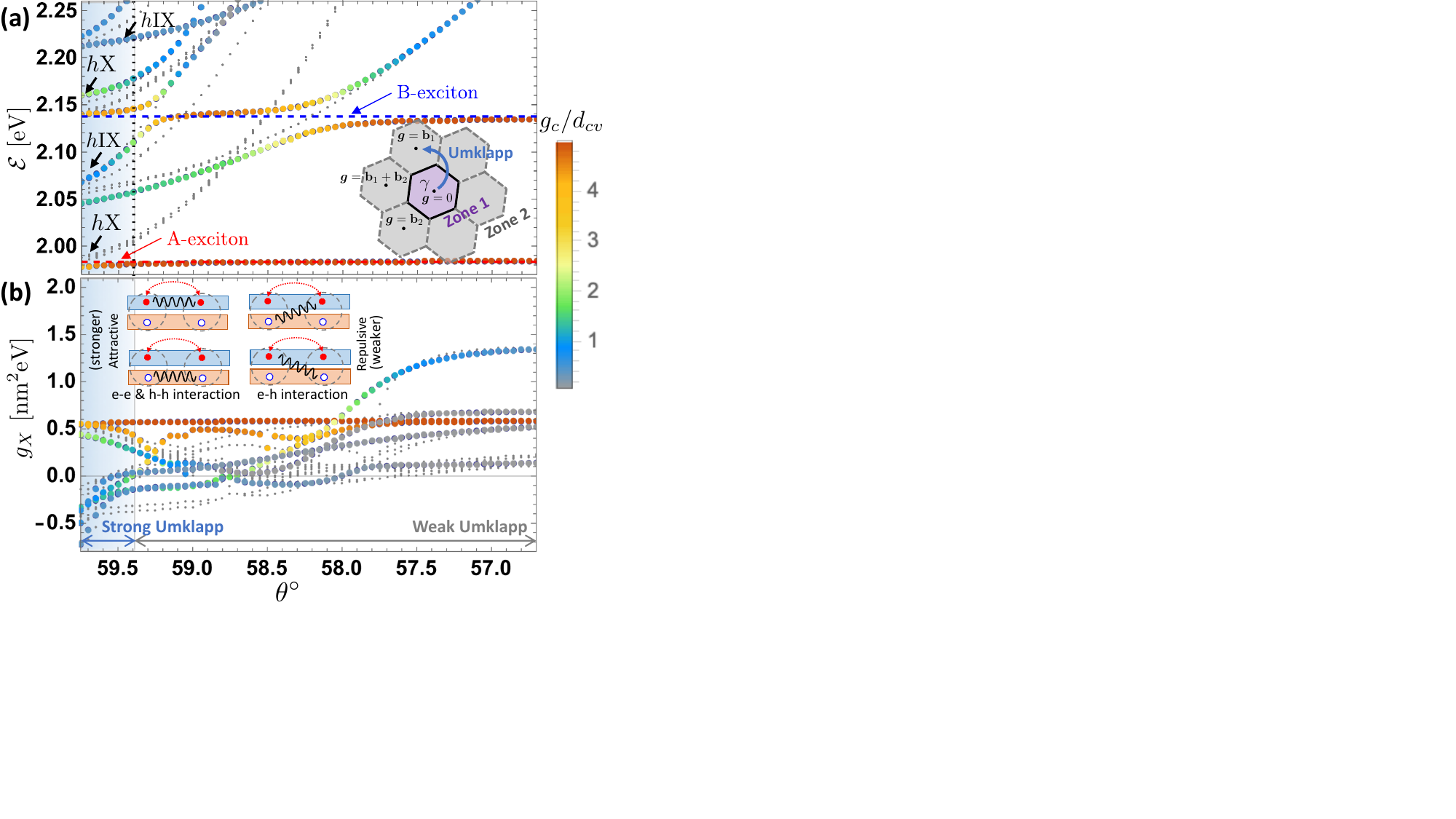}
    \caption{(a) Exciton energy $\mathcal{E}$ ($\sigma=\uparrow$) and twisting angle $\theta$ dependence. The points are color coded by its light-matter coupling strength $g_c$. The hybridized excitons weakly interact with light are shown in gray dots. The red and blue dashed lines are the A and B exciton energy with $\theta=60^{\circ}$. The vertical dashed line is given by $\frac{\Delta\mathbf{K}^2}{2m_v}= t_v$ where the energy separation between folded bands is comparable to interlayer tunneling energy (Inset) the moir\'e exciton spectrum is calculated by considering the Umklapp scattering that hybridizes the exciton in zone 1 (purple area) and zone 2 (gray area). In the weak Umklapp region, the interlayer and intralayer exciton hybridized with each mBZ separately. (b) $\theta$-dependence of the exciton-exciton interaction. (Inset) the interaction (wavy black curves) between interlayer excitons with electron exchange (red dashed curves).}
    \label{fig:MoireXSpec}
\end{figure}

We consider $\bar{\mathbf{Q}}=0$ exciton states being dominant  as large-$\bar{\mathbf{Q}}$ modes are decoupled from light \cite{Wang:RMP90-2018}. Using $\mathcal{H}_{\mathrm{sc}}$ and assuming the exciton in Eq.~\eqref{eqn:X} being a boson described by the operator $\mathcal{X}^{\bar{\alpha}}_{\sigma}(0) \to \hat{x}_{\bar{\alpha}\sigma}$, the photon-cavity coupled system can be written as
\begin{equation}\label{eqn:Hxp}
\mathcal{H}_{\mathrm{xp}}\!=\!\omega_cc^\dagger_0c_0+\sum_{\bar{\alpha}}(\mathcal{E}^{\bar{\alpha}}_{\sigma}\!+\tfrac{1}{L^2}g^{\bar{\alpha}}_X\hat{x}^\dagger_{\bar{\alpha}\sigma}\hat{x}_{\bar{\alpha}\sigma})\hat{x}^\dagger_{\bar{\alpha}\sigma}\hat{x}_{\bar{\alpha}\sigma}+\mathcal{H}_{\mathrm{sc}}
\end{equation}
%
where the X-X interaction is $g^{\bar{\alpha}}_X\approx L^2\Delta^{\bar{\alpha}\bar{\alpha}}$. The exciton-photon interaction is $g_c^{\bar{\alpha}}=\frac{1}{2}\sum_{\alpha\ell}\Omega^\ell_{\alpha}\bar{C}^{\bar{\alpha},\ell\ell}_{\alpha\sigma}$, where $\Omega^\ell_{\alpha}=2\sum_{\mathbf{k}}d^\ell_{cv}\phi^{\ell\ell}_\alpha(\mathbf{k})$ is the Rabi splitting of the exciton in $\ell$-th layer and state $\alpha$. 
\begin{figure*}
    \centering
    \includegraphics[width=6.8in]{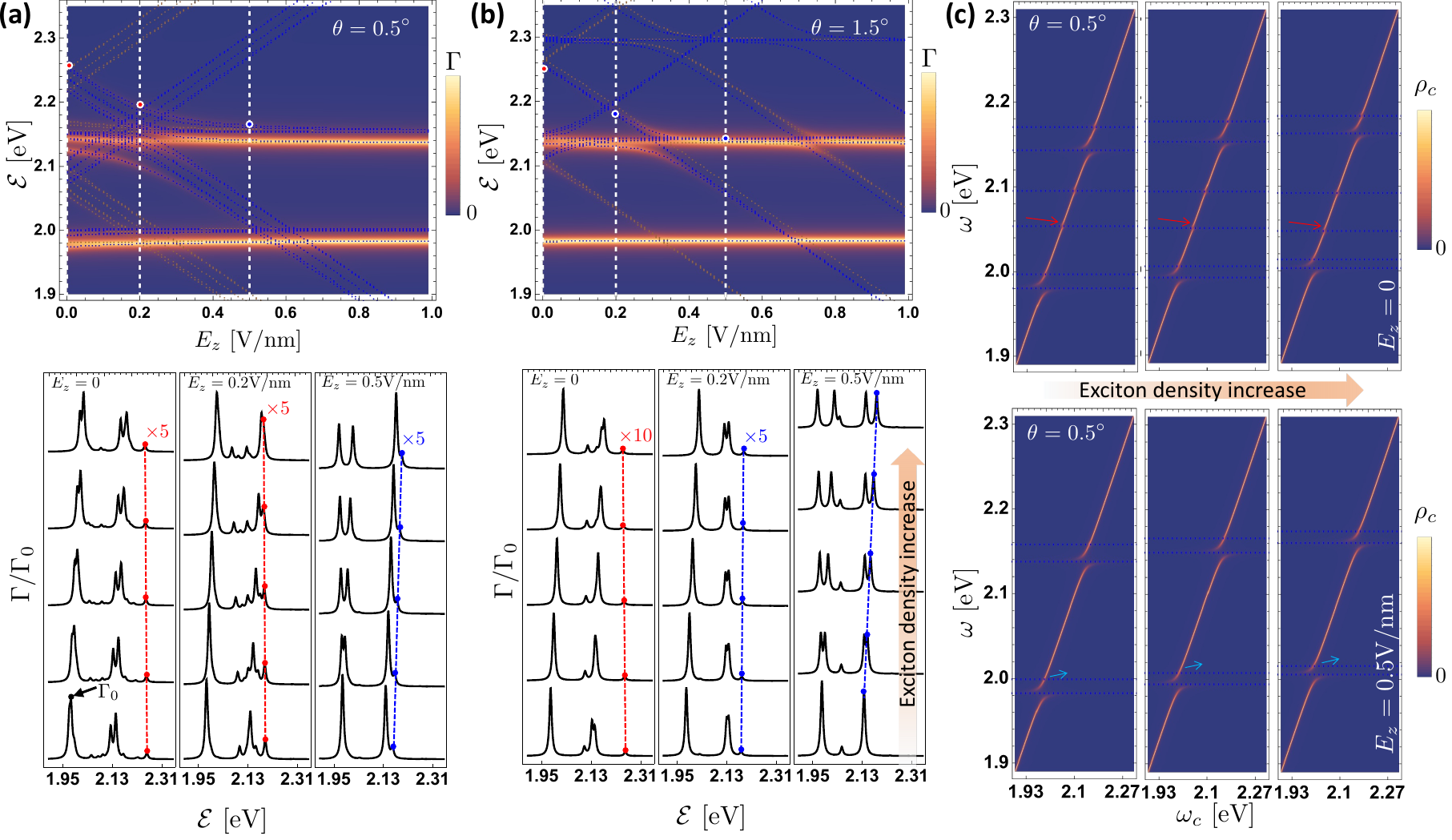}
    \caption{Upper panel is $E_z$-field dependence of the \emph{hybridized} moir\'e exciton absorption spectrum with twisting angles (a) $\theta=0.5^\circ$ and (b) $\theta=1.5^\circ$ (normalized by $\Gamma_0$). The blue ($\sigma=\uparrow$) and red ($\sigma=\downarrow$) dashed curves highlight the moir\'e excitonic modes. Lower panel shows the nonlinear spectrum of the power-dependent absorption with $E_z=0, 0.2, 0.5$~V/nm indicated by the cut (white dashed lines) in the upper panel. Red (blue) dashed lines illustrate the red (blue) shift of the \emph{hybridized} interlayer exciton. (c) Nonlinear polariton spectrum for $\theta=0.5^\circ$ with $\sigma=\uparrow$. In this plot, we let $\kappa=1$~meV and $\gamma_{\bar{\alpha}}=3$~meV.}
    \label{fig:MoireXEz}
\end{figure*}

\emph{Result and discussion.---}To demonstrate the tunability of moir\'e polaritons described by Eq.~\eqref{eqn:Hxp}, we investigate the band hybridization effects for a MoS$_2$ bilayer marginally twisted from anti-parallel stacking/H-type ($\theta\approx 60^\circ$) encapsulated by hexagonal boron nitride ($\epsilon=4$). We adopt the band parameters from Ref.~\cite{Kormanyos:2DMat2-2015} and use interlayer hoppings of $t_c(\mathbf{K}_\ell)=2.1$~meV and $t_v(\mathbf{K}_\ell)=14.5$~meV \cite{Wang:PRB95-2017}. Using the Gaussian basis function expansion~\cite{Ceferino:PRB101-2020,Song:arXiv04-2022,Song:arXiv07-2022}, we solve Eqs.~\eqref{eqn:X-eqn} and \eqref{eqn:hX-eqn} by keeping only the 1$s$ exciton, and plot the $\theta$-dependence of exciton energies $\mathcal{E}^{\bar{\alpha}}_\sigma$ in Fig.~\ref{fig:MoireXSpec}a. The corresponding strength of nonlinearity $g^{\bar{\alpha}}_X$ is shown in Fig.~\ref{fig:MoireXSpec}b. In the plots each point corresponds to moir\'e excitons with an associated index $\bar{\alpha}$, and their light-matter coupling strength ($g_c^{\bar{\alpha}}$) is color-coded. In Fig.~\ref{fig:MoireXSpec}a, we can see two lower branches of bright excitonic modes that correspond to the \emph{hybridized} intralayer exciton hX ($\mathcal{E}\approx1.98$~eV) and \emph{hybridized} interlayer exciton hIX ($\mathcal{E}\approx2.05$~eV). These results match previous calculations for single-particle properties discussed in Ref.~\cite{Ruiz-Tijerina:PRB99-2019}. 

In our analysis, we focus on the nonlinear response and X-X interaction strength shown in Fig.~\ref{fig:MoireXSpec}b. In the plot we keep only the dominant contributions from Eq.~\eqref{eqn:Deltamn} with $\bm{g}'=\bm{g}$ for simplicity, since the Coulomb scattering processes with nonzero momentum transfer between the excitons are rapidly suppressed as $\theta\gtrsim0.5^\circ$. First of all, we find an enhancement in the nonlinear interaction at the small-$\theta$ regime indicating the strong influence of the band hybridization due to the Umklapp processes. We find strong attractive nonlinearity for hIX while the nonlinear interaction for hX remains repulsive. Intriguingly, these results correlate with measurements in MoS$_2$ bilayers, where redshifts were observed \cite{Charalambos:NatCommun14-2023}. The emergence of attractive nonlinearity may be understood as a manifestation of strong anisotropic nature of 2D bilayers. This leads to a weaker interlayer e-h interaction as compared to the intralayer e-e and h-h interaction (Fig.~\ref{fig:MoireXSpec}b, inset). Hence, the total nonlinearity from each of these interacting channels does not cancel leading to stronger attractive nonlinearity. This is different from the monolayer exciton where large cancellations \cite{Song:arXiv04-2022,Song:arXiv07-2022,Shahnazaryan2017} between these channels result in a weaker repulsive nonlinearity ($\sim0.5$~nm$^2$eV). This anisotropic electronic property is unique for 2D materials that enable the realization of attractive interaction between dipolar excitons. 

We also note that the strong attractive X-X interaction between hIX in the small $\theta$ regime comes from the exchange scattering involving the excitonic modes in Zone 2 (Fig.~\ref{fig:MoireXSpec}a, inset). For instance, in the scattering processes with $\bm{g}\neq\tilde{\bm{g}}$ (Fig.~\ref{fig:MoireX}d), the second term in the last line in Eq.\eqref{eqn:Deltamn} (all particle exchange process, repulsive) is strongly suppressed due to large momentum transfer ($\mathbf{q}=\bm{g}-\tilde{\bm{g}}$), while the exchange interaction $V_{\nu\tilde{\nu}}^{\nu'\tilde{\nu}'}$ and $V_{\nu\tilde{\nu}}^{\tilde{\nu}'\nu'}$ remain large. 
This makes a striking difference between moir\'e bilayer and untwisted bilayer ($\theta=0$ and $60^\circ$) where the scattering processes only take place within Zone 1 with $\bm{g}=\tilde{\bm{g}}=0$. Furthermore, in our calculation, we find weak repulsive direct interaction for hIX in a homobilayer with $\theta\approx0$ (see SM\cite{SM}). This may be expected since a large electrical dipole moment is forbidden due to the approximate inversion symmetry. As a result, the dominant contribution to $g_X$ in Fig.~\ref{fig:MoireXSpec}b comes from the (attractive) exchange interactions. However, this can change for heterobilayers or electrically-biased samples. 

In the presence of electric field $\mathbf{E}_z$, the dipole moment of the interlayer exciton couples to $\mathbf{E}_z$ giving rise to the energy Stark shift. This changes the hybridization content of hX and hIX leading to electrically-tunable optical properties. The Stark effect can be incorporated into Eq.~\eqref{eqn:hX-eqn} by modifying the interlayer exciton energy $E_{\nu\sigma}(\bar{\mathbf{Q}})\to E_{\nu\sigma}(\bar{\mathbf{Q}})+\mathbf{E}_z\cdot\bm{\mu}^{s}$, where the dipole moment $\bm{\mu}^{s}=\pm \mu_z\hat{\mathbf{z}}$ for interlayer exciton with $\mu_z \sim 5$e\AA\ \cite{Leisgang:NatNanotech15-2020,Peimyoo2021}, and $\mu_z=0$ for intralayer exciton. In Figs.~\ref{fig:MoireXEz}a and \ref{fig:MoireXEz}b, we calculate the absorption $\Gamma(\mathcal{E})$. In $\Gamma(\mathcal{E})$, we also take into account the nonlinear energy shift due to the background excitons with density $n_{\bar{\alpha}}=\langle\hat{x}^\dagger_{\bar{\alpha}\sigma}\hat{x}_{\bar{\alpha}\sigma}\rangle/L^2$ (pump-power dependent)\cite{SM}. In Figs.~\ref{fig:MoireXEz}a, two brightest branches independent of $E_z$ correspond to the \emph{hybridized} A and B intralayer excitons. The other dimmer branches with strong response to $E_z$ are the hIXs. Following one hIX branch indicated by the red and blue dots in Figs.~\ref{fig:MoireXEz}a and \ref{fig:MoireXEz}b, we can see that the absorption peaks are enhanced as the hIX approaches the hX branches since hIX gains more intralayer exciton component. Furthermore, we also observe that the hIX nonlinear redshift turns into a blueshift as $E_z$ changes. This is partly due to the repulsive nonlinearity from the intralayer exciton component, and another contribution is coming from the polarization of hIX in the high-field regime.
These results in Figs.~\ref{fig:MoireXSpec} and \ref{fig:MoireXEz} demonstrate that the nonlinear optical response of moir\'e material is electrically tunable.

Next, we plot the polaritonic spectrum in Fig.~\ref{fig:MoireXEz}(c) by evaluating the photonic density of states $\rho_c(\omega)$ of Eq.~\eqref{eqn:Hxp}, which is proportional to the cavity transmission \cite{SM}. Tuning the hIX energy closer to the hX energy at $E_z=0.5$~eV/nm, we enhance the light-matter coupling of hIX leading to the larger Rabi splitting. This also changes the hIX attractive nonlinearity (red arrows) at $E_z=0$ to a repulsive nonlinearity (blue arrows) with blueshift at $E_z=0.5$~V/nm. 

\emph{Conclusion.---}We developed a microscopic theory for \emph{hybridized} moir\'e excitons in twisted bilayers.
We revealed that hybridization between layers can enhance significantly the nonlinearity of moir\'e excitons and polaritons, stemming from Umklapp processes for small twisting angles. Intriguingly, we find attractive nonlinear interaction for hIX, which can be tuned into repulsive by applying an external electric field. This makes moir\'e polariton lattices at small twisting angles an excellent platform for studying many-body effects.

Note that in this study we limited ourselves to not-too-large $\theta$ regime, since the hybridization between folded bands in the mini Brillouin zone is weak. On the other hand, we do not investigate $\theta\lesssim0.5^{\circ}$, as this is typically prevented by the lattice reconstruction \cite{Naik:PRL121-2018,Carr:PRB98-2018,Naik:JPhysChemC-2019,Enaldiev:PRL124-2020,Weston:NatNanoTech15-2020} and also requires accounting for more Umklapp scatterings beyond those pictured in Fig.~\ref{fig:MoireXSpec}a. We also stress that the nonlinear properties in our study are valid for low density regimes, and studying higher-order correction is an interesting  avenue for future research. This will allow describing a crossover to the strongly correlated regime close to the metal-insulator transition \cite{Blundo:NatCommun15-2024,Gao:NatCommun15-2024,Huang:PRB107-2023,Park:NatPhys19-2023}. 

\begin{acknowledgements}
\emph{Acknowledgments.}---We acknowledge the support from UK EPSRC Awards No. EP/X017222/1. K. W. S. is supported by Xiamen University Malaysia Research Fund (Grant No. XMUMRF/2025-C15/IPHY/0005).
\end{acknowledgements}

\bibliography{MoireX2}

\clearpage

\setcounter{table}{0}
\setcounter{equation}{0}
\setcounter{figure}{0}
\renewcommand{\theequation}{S\arabic{equation}}
\renewcommand{\thefigure}{S\arabic{figure}}

\begin{widetext}

\section*{Supplemental Material}

\section{Interlayer and intralayer exciton transition matrix element}

The interlayer electron and hole hopping in Eq. \eqref{eqn:Ht} leads to the hybridization of excitons forming a mixture of  intralayer and interlayer exciton as
\begin{equation}\label{eqn:X-xbasis}
\mathcal{X}^{\dagger}_{\sigma}=
\sum_{s\alpha}\sum_{\mathbf{Q}}C^{s}_{\alpha\sigma}(\mathbf{Q})X^{s,\dagger}_{\alpha\sigma}(\mathbf{Q}).
\end{equation}
To find the \emph{hybridized} exciton states, we solve for the eigenstate as $(\mathcal{H}_0 + \mathcal{H}_{\mathrm{int}}+\mathcal{H}_t)\mathcal{X}^{\dagger}_{\sigma}|0\rangle=\mathcal{E}_{\sigma}\mathcal{X}^{\dagger}_{\sigma}|0\rangle$.
This yields the eigenvalue problem as
\begin{equation}
\label{eqn:hybridize-eqn}
    (E^{s}_{\alpha\sigma}(\mathbf{Q})- \mathcal{E}_{\sigma})C^{s}_{\alpha\sigma}(\mathbf{Q})\!=\!\!\!\sum_{s'\alpha'\mathbf{Q}'}\!\!\!W^{s's}_{\alpha'\alpha}(\mathbf{Q}',\mathbf{Q}) C^{s'}_{\alpha'\sigma}(\mathbf{Q}')
\end{equation}
with $\mathcal{E}_\sigma$ being the \emph{hybridized} exciton energy and the transition matrix elements 
\begin{equation}
    W^{s's}_{\alpha'\alpha}(\mathbf{Q}',\mathbf{Q})=\langle 0|X^{s'}_{\alpha'\sigma}(\mathbf{Q}')\mathcal{H}_tX^{s\dagger}_{\alpha\sigma}(\mathbf{Q})|0\rangle.\label{eqn:hyb_M}
\end{equation} 
This determines how the interlayer exciton and intralayer exciton are hybridized. Solving Eq. \eqref{eqn:hybridize-eqn}, we obtain the hybridized exciton in Eq. \eqref{eqn:X-xbasis} with energy $\mathcal{E}_\sigma$.

Using Eqs. \eqref{eqn:Ht} and \eqref{eqn:X}, this gives the transition amplitude
\begin{align}\label{eqn:Wss}
   W^{s's}_{\alpha'\alpha}(\mathbf{Q}',\mathbf{Q})
   =&\sum_{\mathbf{k}}\bar{\phi}_{\alpha'\sigma}^{s'}(\mathbf{k}+\zeta^v_{s'}\mathbf{Q}')T^c_{\ell_c'\ell_c}(\mathbf{k}+\mathbf{Q}'+\delta\mathbf{K},\mathbf{k}+\mathbf{Q}+\delta\mathbf{K})\delta_{\ell_v\ell_v'}\phi_{\alpha\sigma}^{s}(\mathbf{k}+\zeta^v_{s}\mathbf{Q})\notag\\
&-\sum_{\mathbf{k}}\bar{\phi}_{\alpha\sigma}^{s'}(\mathbf{k}-\zeta^c_{s'}\mathbf{Q}')T^v_{\ell_v'\ell_v}(\mathbf{k}-\mathbf{Q}+\delta\mathbf{K},\mathbf{k}-\mathbf{Q}'+\delta\mathbf{K})\delta_{\ell_c'\ell_c}\phi_{\alpha\sigma}^{s}(\mathbf{k}-\zeta^c_{s}\mathbf{Q}),
\end{align}
where $\delta\mathbf{K}=\zeta^v_{s}\mathbf{K}_{\ell_c}+\zeta^c_{s}\mathbf{K}_{\ell_v}$ and the electron/hole mass fraction are $\zeta^{c,v}_{s}=m^{c,v}_{\ell_{c,v}}/M_s$ of $s$-type exciton with total mass $M_s=m^c_{\ell_c}+m^v_{\ell_v}$.

To construct the \emph{hybridized} exciton equation in Eq. \eqref{eqn:hybridize-eqn}, we approximate hopping constant
\cite{Wang:PRB95-2017,Ruiz-Tijerina:PRB99-2019} as in Eq. \eqref{eqn:Tc_Tv}
\begin{align}
    T^{c,v}_{\ell'\ell}(\mathbf{k}',\mathbf{k})=&\sum_{\mathbf{G}_{\ell}\mathbf{G}_{\ell'}}t^{c,v}(\mathbf{K}_\ell+\mathbf{G}_{\ell})\delta_{\mathbf{k}'-\mathbf{k},\mathbf{G}_{\ell'}-\mathbf{G}_{\ell}}\tau^x_{\ell\ell'},
\end{align}
where $\mathbf{G}_{\ell}$ is the reciprocal lattice vector in layer $\ell$. Here, we omit the relative sliding (with vector $\mathbf{r}_0$) between two layers. We note that the momenta are measured from the $\Gamma$-point. 

Expanding the interlayer hopping in the vicinity of valleys, we further approximate Eq. \eqref{eqn:Tc_Tv} by keeping only the $\mathbf{G}_\ell$-sum as follows 
\begin{equation}
    \mathbf{G}_{\ell}=\mathbf{G}_{\ell}^{(\eta)}=(-1)^\eta(C^\eta_3-C^0_3)\mathbf{K}_{\ell}, \quad (\eta=1,2,3)
\end{equation}
which connects the three equivalent valleys within the first Brillouin zone. Here, $C_3$ is the $2\pi/3$-rotation operator for the in-plane vector. Because of the $C_3$ crystal symmetry, we have
\begin{align}
    t^c(\mathbf{K}_\ell+\mathbf{G}^{(\eta)}_\ell)=&t^c,\notag\\
    t^v(\mathbf{K}_\ell+\mathbf{G}^{(\eta)}_\ell)=&t^v\begin{cases}
        1, & \text{(P stacking)}\\
        \mathrm{e}^{-i\frac{4}{3}\pi \eta},& \text{(AP stacking)}
    \end{cases}
\end{align}
for $\eta=0,1,2$. 
Keeping only the interlayer scattering with $|\mathbf{G}_{\ell}-\mathbf{G}_{\ell'}|\leq|\mathbf{b}_{1}|$, this approximation leads to
\begin{align}
T^{c}_{\ell'\ell}(\mathbf{k}',\mathbf{k})=t^{c}\sum_{\eta=0}^{2}\delta_{\mathbf{k}'-\mathbf{k},\mathbf{b}_\eta}, \quad T^{v}_{\ell'\ell}(\mathbf{k}',\mathbf{k})=t^{v}\sum_{\eta=0}^{2}\delta_{\mathbf{k}'-\mathbf{k},\mathbf{b}_\eta}\begin{cases}
        1, & \text{(P stacking)}\\
        \mathrm{e}^{-i\frac{4}{3}\pi \eta},& \text{(AP stacking)}
    \end{cases}
\end{align}
with $\mathbf{b}_0=0$. We note that this result has been obtained in Ref. \cite{Ruiz-Tijerina:PRB99-2019}. In their notation, the momentum $\mathbf{k}$ and $\mathbf{k}'$ are measured from the valleys $\mathbf{K}_{1,2}$, while in our paper we measured the momentum from the $\Gamma$ point. With this hopping constant, this reduces Eq. \eqref{eqn:Wss} (P stacking) to
\begin{align}\label{eqn:W-M}
    W^{s's}_{\alpha'\alpha}(\mathbf{Q},\mathbf{Q}')=&\sum_{\mathbf{k}}\Big[t_c\bar{\phi}_{\alpha'\sigma}^{s'}[\mathbf{k}\!+\!\zeta^{v}_{s'}(\bar{\mathbf{Q}}'\!+\!\bm{g}')]\phi_{\alpha\sigma}^{s}[\mathbf{k}\!+\!\zeta^v_{s}(\bar{\mathbf{Q}}\!+\!\bm{g})]\delta_{\ell_v'\ell_v}\tau^{x}_{\ell_c'\ell_c}\notag\\
    &
    -t_v\bar{\phi}_{\alpha'\sigma}^{s'}[\mathbf{k}\!-\!\zeta^{c}_{s'}(\bar{\mathbf{Q}}'\!+\!\bm{g}')]\phi_{\alpha\sigma}^{s}[\mathbf{k}\!-\!\zeta^c_{s}(\bar{\mathbf{Q}}\!+\!\bm{g})]\delta_{\ell_c\ell_c'}\tau^{x}_{\ell_v'\ell_v}\Big]\sum_{\eta=0}^2\delta_{\bm{g}'-\bm{g},\mathbf{b}_{\eta}-(\bar{\mathbf{Q}}'-\bar{\mathbf{Q}})},
\end{align}
where $\mathbf{Q}=\bar{\mathbf{Q}}+\bm{g}$ with $\bar{\mathbf{Q}}$ being the momentum restricted in the mini Brillouin zone of the moir\'e superlattice (see Fig.~\ref{fig:MoireXSpec}a). 
To further simplify the above, we observe that $\bm{g}=i\mathbf{b}_1+j\mathbf{b}_2$ is also reciprocal lattice vector of a moir\'e lattice. This implies that Eq.~\eqref{eqn:W-M} is zero if $\bar{\mathbf{Q}}'\neq\bar{\mathbf{Q}}$. This gives 
\begin{align}\label{eqn:Wss-delta}
	W^{s's}_{\alpha'\alpha}(\mathbf{Q},\mathbf{Q}')=w_{\nu'\nu}(\bar{\mathbf{Q}})\delta_{\bar{\mathbf{Q}}',\bar{\mathbf{Q}}}.
\end{align}
This result is expected since the delta function $\delta_{\bar{\mathbf{Q}}',\bar{\mathbf{Q}}}$ reflects the translational symmetry of the moir\'e superlattice. Therefore, the reduced transition matrix element is 
\begin{align}\label{eqn:w-amp}
    w_{\nu'\nu}(\bar{\mathbf{Q}})=&(t_cM^{v}_{\nu'\nu}(\bar{\mathbf{Q}})\delta_{\ell_v'\ell_v}\tau^{x}_{\ell_c'\ell_c}
    -t_vM^{c}_{\nu'\nu}(\bar{\mathbf{Q}})\delta_{\ell_c'\ell_c}\tau^{x}_{\ell_v'\ell_v})\sum_{\eta=0}^2\delta_{\bm{g}'-\bm{g},\mathbf{b}_{\eta}}.
\end{align}
The transition matrix element between interlayer-intralayer exciton is 
\begin{align*}
    M^{c}_{\nu'\nu}(\bar{\mathbf{Q}})\!&=\!\!\sum_{\mathbf{k}}\bar{\phi}_{\alpha'\sigma}^{s'}[\mathbf{k}\!-\!\zeta^{c}_{s'}(\bar{\mathbf{Q}}\!+\!\bm{g}')]\phi_{\alpha\sigma}^{s}[\mathbf{k}\!-\!\zeta^c_{s}(\bar{\mathbf{Q}}\!+\!\bm{g})],\\
    M^{v}_{\nu'\nu}(\bar{\mathbf{Q}})\!&=\!\!\sum_{\mathbf{k}}\bar{\phi}_{\alpha'\sigma}^{s'}[\mathbf{k}\!+\!\zeta^{v}_{s'}(\bar{\mathbf{Q}}\!+\!\bm{g}')]\phi_{\alpha\sigma}^{s}[\mathbf{k}\!+\!\zeta^v_{s}(\bar{\mathbf{Q}}\!+\!\bm{g})].
\end{align*}
The summation of the delta functions in Eq.~\eqref{eqn:w-amp} correspond to the mixing of the seven mBZs shown in the inset of Fig.~\ref{fig:MoireXSpec}a. One may include more mixing beyond these mBZ. However, this involves the electrons with higher-energy dispersion and their effects will be suppressed. 

Due to the conservation of $\bar{\mathbf{Q}}$ in Eq.~\eqref{eqn:Wss-delta}, each different $\bar{\mathbf{Q}}$ modes in Eq.~\eqref{eqn:hybridize-eqn} do not mix together and break into independent sectors. This yields Eq.~\eqref{eqn:hX-eqn},
\begin{equation}
        [E_{\nu\sigma}(\bar{\mathbf{Q}})-\mathcal{E}^{\bar{\alpha}}_{\sigma}(\bar{\mathbf{Q}})]C^{\bar{\alpha}}_{\nu\sigma}(\bar{\mathbf{Q}})
\!=\!
\sum_{\nu'}w_{\nu'\nu}(\bar{\mathbf{Q}})C^{\bar{\alpha}}_{\nu'\sigma}(\bar{\mathbf{Q}}),
\end{equation}
where we used the shorthand index $\nu=(s,\alpha,\bm{g})$ to represent exciton species $s$, exciton state $\alpha$, and total momentum $\bm{g}$ to lighten our notation. Also, we let $E_{\nu\sigma}(\bar{\mathbf{Q}})=E^s_{\alpha\sigma}(\bar{\mathbf{Q}}+\bm{g})$ and $C^{\bar{\alpha}}_{\nu\sigma}(\bar{\mathbf{Q}})=C^{s}_{\alpha\sigma}(\bar{\mathbf{Q}}+\bm{g})$. Here, we label the \emph{hybridized} exciton energy eigenstates by index $\bar{\alpha}$ where we use index with a bar to indicate the quantum number of the hybridized states.

\section{X-X interaction}
In calculating the X-X interactions, we focus on the $\bar{\mathbf{Q}}=0$ case. The procedure follows the composite boson treatment described in Ref.~\cite{Combescot:PhysRep5-2008}. The interaction strength is characterized by the energy of the two excitonic states 
\begin{equation}
    \Omega^{\bar{\beta}\bar{\alpha}}_{\sigma}=\langle0|\mathcal{X}^{\bar{\alpha}}_{\sigma}\mathcal{X}^{\bar{\beta}}_{\sigma}\mathcal{H}\mathcal{X}^{\bar{\beta},\dagger}_{\sigma}\mathcal{X}^{\bar{\alpha},\dagger}_{\sigma}|0\rangle .
\end{equation}
To evaluate $\Omega^{\bar{\beta}\bar{\alpha}}_{\sigma}$, we substitute Eq.~\eqref{eqn:X-mix}, and use the commutation relation  
\begin{align*}
    [\rho_{\ell\mathbf{q}},X^{s,\dagger}_{\alpha\sigma}(\mathbf{Q})]
    =
    \sum_{\mathbf{k}_c\mathbf{k}_v}\sum_{\tilde{\alpha}\tilde{\mathbf{Q}}}F^{\ell,s}_{\alpha\mathbf{Q}\sigma}(\mathbf{k}_c\mathbf{k}_v;\mathbf{q})\bar{\Phi}^{s}_{\tilde{\alpha}\tilde{\mathbf{Q}},\sigma}(\mathbf{k}_c,\mathbf{k}_v)X^{s,\dagger}_{\tilde{\alpha}\sigma}(\tilde{\mathbf{Q}}),
\end{align*}
where $\bar{\Phi}^{s}_{\alpha\mathbf{Q},\sigma}(\mathbf{k}_c,\mathbf{k}_v)=\delta_{\mathbf{k_c-\mathbf{k}_v},\mathbf{Q}}\phi^{s}_{\alpha\sigma}(m^v_{\ell_v}(\mathbf{k}_c-\mathbf{K}_{\ell_c})+m^c_{\ell_c}(\mathbf{k}_v-\mathbf{K}_{\ell_v})]/(m^c_{\ell_c}+m^v_{\ell_v}))$ is the exciton wavefunction in Eq.~\eqref{eqn:X}, the density operator  $\rho_{\ell\mathbf{q}}=\sum_{\mathbf{k}\sigma}(a^{\ell\dagger}_{ \mathbf{k}+\mathbf{q},\sigma}a^{\ell}_{\mathbf{k} \sigma}+b^{\ell\dagger}_{ \mathbf{k}+\mathbf{q},\sigma}b^{\ell}_{\mathbf{k} \sigma})$, and
\begin{equation*}
    F^{\ell,s}_{\alpha\mathbf{Q}\sigma}(\mathbf{k}_c\mathbf{k}_v;\mathbf{q})=[\delta_{\ell_c\ell}\Phi^{s}_{\alpha\mathbf{Q}\sigma}(\mathbf{k}_c-\mathbf{q},\mathbf{k}_v)-\delta_{\ell_v\ell}\Phi^{s}_{\alpha\mathbf{Q}\sigma}(\mathbf{k}_c,\mathbf{k}_v+\mathbf{q})] .
\end{equation*}
In the derivation of the above commutation relation, we have used the completeness relation for the exciton wavefunction 
\begin{align}
\sum_{\alpha\mathbf{Q}}\bar{\Phi}^{\ell_c\ell_v}_{\alpha\mathbf{Q},\sigma}(\mathbf{k}_c',\mathbf{k}_v')\Phi^{\ell_c\ell_v}_{\alpha\mathbf{Q},\sigma}(\mathbf{k}_c,\mathbf{k}_v)=\delta_{\mathbf{k}_c\mathbf{k}_c'}\delta_{\mathbf{k}_v\mathbf{k}_v'},\quad
\sum_{\alpha\mathbf{Q}}\bar{\Phi}^{\ell_c\ell_v}_{\alpha\mathbf{Q},\sigma}(\mathbf{k}_c,\mathbf{k}_v)X^{\ell_c\ell_v}_{\alpha\mathbf{Q},\sigma}=a^{\ell_c\dagger}_{\mathbf{k}_c\sigma}b^{\ell_v}_{\mathbf{k}_v\sigma} .
\end{align}

Using the Wannier equation in Eq.~\eqref{eqn:X-eqn}, the exciton hybridization equation in Eq.~\eqref{eqn:hX-eqn}, and the above completeness relation, we arrive at
\begin{equation}
\Omega^{mn}_{\sigma}=\mathcal{E}^{\bar{\beta}}_{\sigma}+\mathcal{E}^{\bar{\alpha}}_{\sigma}+\Delta^{mn}_{\sigma},
\end{equation}
where the interacting term is
\begin{align*}
\Delta^{\bar{\beta \bar{\alpha}}}_{\sigma}=
&\sum_{\nu\tilde{\nu}}\sum_{\nu'\tilde{\nu}'}\bar{C}^{\bar{\beta}}_{\tilde{\nu}'\sigma}\bar{C}^{\bar{\alpha}}_{\nu',\sigma}C^{\bar{\beta}}_{\tilde{\nu},\sigma}C^{\bar{\alpha}}_{\nu\sigma}\sum_{\ell\ell',\mathbf{q}}\frac{v_{\ell\ell'}(\mathbf{q})}{L^2}
\sum_{\mathbf{k}_c\mathbf{k}_v}\sum_{\tilde{\mathbf{k}}_c\tilde{\mathbf{k}}_v}F^{\ell,s}_{\alpha\bm{g}\sigma}(\mathbf{k}_c\mathbf{k}_v;\mathbf{q})F^{\ell',\tilde{s}}_{\tilde{\alpha}\tilde{\bm{g}}\sigma}(\tilde{\mathbf{k}}_c\tilde{\mathbf{k}}_v;-\mathbf{q})\notag\\
&\sum_{\alpha''\tilde{\alpha}''}\sum_{\bm{g}''\tilde{\bm{g}}''}\bar{\Phi}^{\tilde{s}}_{\tilde{\alpha}''\tilde{\bm{g}}''\sigma}(\tilde{\mathbf{k}}_c,\tilde{\mathbf{k}}_v)\bar{\Phi}^{s}_{\alpha''\bm{g}''\sigma}(\mathbf{k}_c,\mathbf{k}_v)\langle0|X_{\tilde{\nu}'\sigma}(0)X_{\nu'\sigma}(0)X^{s,\dagger}_{\alpha''\bm{g}''\sigma}(0)X^{\tilde{s},\dagger}_{\tilde{\alpha}''\tilde{\bm{g}}''\sigma}(0)|0\rangle .
\end{align*}
Here, we recall that the $X_{\nu'\sigma}(0)=X^{s'}_{\alpha'\bm{g}'\sigma}(0)$ and $X_{\tilde{\nu}'\sigma}(0)=X^{\tilde{s}'}_{\tilde{\alpha}'\tilde{\bm{g}}'\sigma}(0)$.

To proceed further, we use
\begin{equation}\label{eqn:[X,X+]}
    [ X^{s'}_{\alpha'\sigma}(\mathbf{Q}'),X^{s,\dagger}_{\alpha\sigma}(\mathbf{Q})]=\delta_{\alpha\alpha'}\delta_{ss'}\delta_{\mathbf{Q},\mathbf{Q}'}-D^{s's}_{\alpha'\alpha}(\mathbf{Q}',\mathbf{Q})
\end{equation} 
with the non-bosonicity
\begin{align}
    D^{s's}_{\alpha'\alpha}(\mathbf{Q}',\mathbf{Q})=&\sum_{\mathbf{k}_c\mathbf{k}_c'}\sum_{\mathbf{k}_v\mathbf{k}_v'}\Phi^{s}_{\alpha\mathbf{Q}}(\mathbf{k}_c,\mathbf{k}_v)\bar{\Phi}^{s'}_{\alpha'\mathbf{Q}'}(\mathbf{k}_c',\mathbf{k}_v')\Big\{
  \delta_{\ell_v\ell_v'}\delta_{\mathbf{k}_v\mathbf{k}_v'}a^{\ell_c\dagger}_{\mathbf{k}_c}a^{\ell_c'}_{\mathbf{k}_c'}+\delta_{\ell_c\ell_c'}\delta_{\mathbf{k}_c'\mathbf{k}_c}b^{\ell_v}_{\mathbf{k}_v}b^{\ell_v'\dagger}_{\mathbf{k}_v'}
  \Big\} .
\end{align}
Letting $D^{s's}_{\nu'\nu}|0\rangle=0$, we reduce the calculation of $\Delta^{mn}_{\sigma}$ to the following
\begin{align}
    \langle0|X^{\tilde{s}'}_{\tilde{\alpha}'\sigma}(\tilde{\mathbf{Q}}')X^{s'}_{\alpha'\sigma}(\mathbf{Q}')X^{s,\dagger}_{\alpha\sigma}(\mathbf{Q})X^{\tilde{s},\dagger}_{\tilde{\alpha}\sigma}(\tilde{\mathbf{Q}})|0\rangle=&\delta_{\alpha\alpha'}\delta_{\tilde{\alpha}\tilde{\alpha}'}\delta_{\mathbf{Q}\mathbf{Q}'}\delta_{\tilde{\mathbf{Q}}\tilde{\mathbf{Q}}'}\delta_{ss'}\delta_{\tilde{s}\tilde{s}'}+\delta_{\alpha\tilde{\alpha}'}\delta_{\tilde{\alpha}\alpha'}\delta_{\mathbf{Q}\tilde{\mathbf{Q}}'}\delta_{\tilde{\mathbf{Q}}\mathbf{Q}'}\delta_{s\tilde{s}'}\delta_{\tilde{s}s'}
    \notag\\
    &-\langle0|X^{\tilde{s}'}_{\tilde{\alpha}'\sigma}(\tilde{\mathbf{Q}}')D^{s's}_{\alpha'\alpha}(\mathbf{Q}',\mathbf{Q})X^{\tilde{s},\dagger}_{\tilde{\alpha}\sigma}(\tilde{\mathbf{Q}})|0\rangle .
\end{align}
Using $D^{s's}_{\alpha'\alpha}(\mathbf{Q}',\mathbf{Q})|0\rangle=0$, the commutation relation Eq. \eqref{eqn:[X,X+]}, and
\begin{align}
[D^{s's}_{\alpha'\alpha}(\mathbf{Q}',\mathbf{Q}),X^{\tilde{s},\dagger}_{\tilde{\alpha}\sigma}(\tilde{\mathbf{Q}})]=&\sum_{\mathbf{k}_c\mathbf{k}_v}\sum_{\tilde{\mathbf{k}}_v\tilde{\mathbf{k}}_c}\Phi^{s}_{\alpha\mathbf{Q}\sigma}(\mathbf{k}_c,\mathbf{k}_v)\Phi^{\tilde{s}}_{\tilde{\alpha}\tilde{\mathbf{Q}}\sigma}(\tilde{\mathbf{k}}_c,\tilde{\mathbf{k}}_v)\Big\{
  \delta_{\ell_v\ell_v'}\delta_{\ell_c'\tilde{\ell}_c}\bar{\Phi}^{s'}_{\alpha'\mathbf{Q}'\sigma}(\tilde{\mathbf{k}}_c,\mathbf{k}_v)\Phi^{\ell_c\tilde{\ell}_v}_{\beta',\mathbf{k}_c-\tilde{\mathbf{k}}_v}(\mathbf{k}_c,\tilde{\mathbf{k}}_v)X^{\ell_c\tilde{\ell}_v,\dagger}_{\beta',\mathbf{k}_c-\tilde{\mathbf{k}}_v}\notag\\
    &
  +
\delta_{\ell_c\ell_c'}\delta_{\ell_v'\tilde{\ell}_v}\bar{\Phi}^{s'}_{\alpha'\mathbf{Q}'\sigma}(\mathbf{k}_c,\tilde{\mathbf{k}}_v)\Phi^{\tilde{\ell}_c\ell_v}_{\beta',\tilde{\mathbf{k}}_c-\mathbf{k}_v}(\tilde{\mathbf{k}}_c,\mathbf{k}_v)X^{\tilde{\ell}_c\ell_v,\dagger}_{\beta',\tilde{\mathbf{k}}_c-\mathbf{k}_v}
  \Big\} .
\end{align}
Therefore, we finally arrive at Eq. \eqref{eqn:Deltamn}.

\section{Momentum integration in direct and exchange interactions}

To numerically evaluate Eq. \eqref{eqn:Deltamn}, we expand the exciton wavefunction in the direct interaction $U^{\nu'\tilde{\nu}'}_{\nu\tilde{\nu}}$ and exchange interaction $V^{\nu'\tilde{\nu}'}_{\nu\tilde{\nu}}$ into Gaussian basis function $\varphi_{n}(k\lambda)=\frac{\lambda}{n!2^n\sqrt{\pi}}\exp(-\frac{1}{2}\lambda^2k^2)H_n(k\lambda)$ with Hermite polynomial $H_n(x)$ as 
\begin{equation}
	\phi^{s}_{\alpha\sigma}(\mathbf{p})=\sum_{n_xn_y}S^{s}_{\alpha,\bm{n}}\varphi_{n_x}(p_x\lambda)\varphi_{n_y}(p_y\lambda) ,
\end{equation}
where $\bm{n}=(n_x,n_y)$ is a double index to label the basis functions. Here, the expansion coefficients $S^s_{\alpha\bm{n}}$ are obtained by diagonalizing Eq.~\eqref{eqn:X-eqn}. The length parameter for the basis set $\lambda$ is obtained by optimizing eigenenergies in Eq.~\eqref{eqn:X-eqn}. In the following, we omit calculation of direct interaction which is straightforward, but we will focus on the calculation of exchange interaction.

First, we shift $\mathbf{k}\to\mathbf{k}+\mathbf{q}$. We can then separate the $\mathbf{q}$ and $\mathbf{k}$ integration as 
\begin{align}
    V^{\nu'\tilde{\nu}'}_{\nu\tilde{\nu}}=&\frac{1}{L^2}\sum_{\mathbf{q}}\sum_{\bm{n}_1\bm{n}_2\bm{n}_3\bm{n}_4}S^{s}_{\alpha,\bm{n}}S^{\tilde{s}}_{\tilde{\alpha},\tilde{\bm{n}}}S^{s'}_{\alpha',\bm{n}'}S^{\tilde{s}'}_{\tilde{\alpha}',\tilde{\bm{n}}'}\Big[w_{\ell_c\tilde{\ell}_c}(\mathbf{q})I^{cc}_{\vec{n}_x}(q_x)I^{cc}_{\vec{n}_y}(q_y)+w_{\ell_v\tilde{\ell}_v}(\mathbf{q})I^{vv}_{\vec{n}_x}(q_x)I^{vv}_{\vec{n}_y}(q_y)
    \notag\\
    &-w_{\ell_c\tilde{\ell}_v}(\mathbf{q})I^{cv}_{\vec{n}_x}(q_x)I^{cv}_{\vec{n}_y}(q_y)-w_{\ell_v\tilde{\ell}_c}(\mathbf{q})I^{vc}_{\vec{n}_x}(q_x)I^{vc}_{\vec{n}_y}(q_y)\Big]\delta_{\ell_v\ell_v'}\delta_{\ell_c'\tilde{\ell}_c}\delta_{\ell_c\tilde{\ell}_c'}\delta_{\tilde{\ell}_v\tilde{\ell}_v'}.
\label{eqn:V_exp}
\end{align}
We find that the $\mathbf{k}$ integration reduces to evaluating
\begin{align}
    I^{\xi}_{\vec{m}}(q)=\int d^2k\prod_{i=1}^{4}\frac{\sqrt{\lambda_i}}{\sqrt{m_i!2^{m_i}\sqrt{\pi}}}\exp\Big[-\tfrac{1}{2}\lambda_i^2(k-u^\xi_{q,i})^2\Big]H_{m_i}(\lambda_i(k-u^\xi_{q,i})),
\end{align}
where $\xi=cc,cv,vc,vv$, and $\vec{m}=[m_1,\dots,m_4]$, $\vec{\lambda}=[\lambda_1,\dots,\lambda_4]$ and $\vec{u}^{\xi}(q)=[u^{\xi}_{1}(q),\dots,u_{4}^{\xi}(q)]$. 
In Eq. \eqref{eqn:V_exp},
\begin{align}\label{eqn:v^cc}
    \vec{u}^{cc}(q)=\begin{bmatrix}
        q,
        & 
         0,
        & 
        q,
        &
        0
    \end{bmatrix}+\vec{g}, \quad&
    \vec{u}^{vv}(q)=\begin{bmatrix}
        0,
        & 
        q,
        & 
        q,
        &
        0
    \end{bmatrix}+\vec{g},\\
    \vec{u}^{cv}(q)=\begin{bmatrix}
        0,
        & 
        0,
        & 
        0,
        &
        q
    \end{bmatrix}+\vec{g},
    \quad&
    \vec{u}^{vc}(q)=\begin{bmatrix}
        0,
        & 
        0,
        & 
        q,
        &
        0
    \end{bmatrix}+\vec{g}\label{eqn:v^vc}
\end{align}
where $\vec{g}_{x,y}=[0,1,0,1](g_{x,y}-g'_{x,y})+[\zeta^c_sg_{x,y},\zeta_{\tilde{s}}^c\tilde{g}_{x,y},\zeta^c_{s'}g'_{x,y},\zeta_{\tilde{s}'}^c\tilde{g}'_{x,y}]$.
Similarly, the $y$-component can be obtained by the similar fashion.
To proceed further, we rewrite the above into
\begin{align}
    I^\xi_{\vec{m}}(q)
        =&
    \mathrm{e}^{-\frac{1}{2}\vec{\lambda}^2[\bar{u}_2^\xi(q)-(\bar{u}^\xi_1(q))^2]}\int d^2k\mathrm{e}^{-\frac{1}{2}\vec{\lambda}^2k^2}\prod_{i=1}^{4}\frac{\sqrt{\lambda_i}}{\sqrt{m_i!2^{m_i}\sqrt{\pi}}}H_{m_i}(\lambda_i(k+\bar{u}^\xi_1(q)-u_{i}^\xi(q))),
\end{align}
where $\bar{u}_1^\xi(q)=\sum_{i=1}^4\lambda_i^2u_{i}^\xi(q)/\vec{\lambda}^2$, $\bar{u}_2^\xi(q)=\sum_{i=1}^4\lambda_i^2u_{i}^2(q)/\vec{\lambda}^2$ .
We can integrate out $k$. This yields
\begin{align}
    I^\xi_{\vec{m}}(q)
        =&
    \mathrm{e}^{-\frac{1}{2}\vec{\lambda}^2[\bar{u}_2^\xi(q)-(\bar{u}^\xi_1(q))^2]}\Big[\prod_{i=1}^{4}\sum_{r_i=0}^{\lfloor m_i/2\rfloor}\sum_{t_i=0}^{m_i-2r_i}[\bar{u}_1^\xi(q)-v_{i}(q)]^{t_i}\Big]M^{\vec{r},\vec{t}}_{\vec{m}}(\vec{\lambda}) ,
\end{align}
where $\vec{r}=[r_1,\dots,r_4]$, $\vec{t}=[t_1,\dots,t_4]$, and  
\begin{align}
    M^{\vec{r},\vec{t}}_{\vec{m}}(\vec{\lambda})=\Big[\prod_{i=1}^{4}\frac{\sqrt{\lambda_i}}{\sqrt{m_i!2^{m_i}\sqrt{\pi}}}\frac{(-1)^{r_i}m_i!(2\lambda_i)^{m_i-2r_i}}{r_i!t_i!(m_i-2r_i-t_i)!}\Big(\frac{2}{\vec{\lambda}^2}\Big)^{(m_i-2r_i-t_i+1)/2}\Big]\frac{1}{2}\Gamma\Big(\frac{\sum_{i=1}^{4}(m_i-2r_i-t_i)+1}{2}\Big)
\end{align}
Substituting Eqs.~\eqref{eqn:v^cc}-\eqref{eqn:v^vc} into $b$ and $c$ we have
\begin{align}
    \bar{u}^\xi_1(q)-u_i^\xi(q)=\bar{G}_1-g_i+\begin{cases}
        (\bar{\lambda}^2_{1}+\bar{\lambda}^2_{3})q-(\delta_{i1}+\delta_{i3})q,& \text{for } \xi=cc\\
        (\bar{\lambda}^2_{2}+\bar{\lambda}^2_{3})q-(\delta_{i2}+\delta_{i3})q,& \text{for } \xi=vv\\
        \bar{\lambda}^2_{4}q-\delta_{i4}q,& \text{for } \xi=cv\\
        \bar{\lambda}^2_{3}q-\delta_{i3}q,& \text{for } \xi=vc\\
    \end{cases}
\end{align}
with $\bar{\lambda}_i=\lambda_i/|\vec{\lambda}|$  and $\bar{G}_1=\sum_{i=1}^4\bar{\lambda}_i^2g_i$. We note in the above that $\delta_{ij}$ is the Kronecker delta. Also, we have
\begin{align}
    \bar{u}^\xi_2(q)-[\bar{u}^\xi_1(q)]^2=&
    \bar{G}_2-\bar{G}_1^2+\begin{cases}
    2[\bar{\lambda}_1^2(g_1-\bar{G}_1)+\bar{\lambda}_3^2(g_3-\bar{G}_1)]q+[\bar{\lambda}_1^2+\bar{\lambda}_3^2-(\bar{\lambda}_1^2+\bar{\lambda}_3^2)^2]q^2,&\text{for } \xi=cc\\
    2[\bar{\lambda}_2^2(g_2-\bar{G}_1)+\bar{\lambda}_3^2(g_3-\bar{G}_1)]q+[\bar{\lambda}_2^2+\bar{\lambda}_3^2-(\bar{\lambda}_2^2+\bar{\lambda}_3^2)^2]q^2,&\text{for } \xi=vv\\
    2\bar{\lambda}_4^2(g_4-\bar{G}_1)q+[\bar{\lambda}_4^2-(\bar{\lambda}_4^2)^2]q^2,&\text{for } \xi=cv\\
    2\bar{\lambda}_3^2(g_3-\bar{G}_1)q+[\bar{\lambda}_3^2-(\bar{\lambda}_3^2)^2]q^2,&\text{for } \xi=vc
    \end{cases}
\end{align}
where $\bar{G}_2=\sum_{i=1}^4\bar{\lambda}_i^2g_i^2$. Therefore, we arrive at the following expression for the interaction: 
\begin{align}
    V^{\nu'\tilde{\nu}'}_{\nu\tilde{\nu}}=&\frac{1}{L^2}\sum_{\bm{n}_1\bm{n}_2\bm{n}_3\bm{n}_4}S^{s}_{\alpha,\bm{n}}S^{\tilde{s}}_{\tilde{\alpha},\tilde{\bm{n}}}S^{s'}_{\alpha',\bm{n}'}S^{\tilde{s}'}_{\tilde{\alpha}',\tilde{\bm{n}}'}\Big[\prod_{i=1}^{4}\sum_{r_{xi}=0}^{\lfloor n_{xi}/2\rfloor}\sum_{t_{xi}=0}^{n_{xi}-2r_{xi}}\sum_{r_{yi}=0}^{\lfloor n_{yi}/2\rfloor}\sum_{t_{yi}=0}^{n_{yi}-2r_{yi}}\Big]
    \notag\\
    &\Big[\Theta^{cc}_{\vec{t}_x,\vec{t}_y}(\vec{\lambda},\vec{g})+\Theta^{vv}_{\vec{t}_x,\vec{t}_y}(\vec{\lambda},\vec{g})-\Theta^{cv}_{\vec{t}_x,\vec{t}_y}(\vec{\lambda},\vec{g})-\Theta^{vc}_{\vec{t}_x,\vec{t}_y}(\vec{\lambda},\vec{g})\Big]M^{\vec{r}_x\vec{t}_x}_{\vec{n}_x}(\vec{\lambda})M^{\vec{r}_y\vec{t}_y}_{\vec{n}_y}(\vec{\lambda})\delta_{\ell_v\ell_v'}\delta_{\ell_c'\tilde{\ell}_c}\delta_{\ell_c\tilde{\ell}_c'}\delta_{\tilde{\ell}_v\tilde{\ell}_v'}.
\end{align}

Now, we consider that the expression above is $\mathbf{q}$-dependent, and try to integrate $q$ out. The integration in $\mathbf{q}$ generally does not have an analytical form, since the potential may be arbitrary. Therefore, we evaluate it numerically. However, the potential has rotational symmetry. We may reduce the two-dimensional integral into a sum of a fast convergent series. The ultimate integration is the following
\begin{equation}
    \Theta^\xi_{\vec{t}_x,\vec{t}_y}(\vec{\lambda},\vec{g})=\int_0^{\infty} qdq\int_{0}^{2\pi}d\theta  v^{\xi}(\mathbf{q})\mathrm{e}^{-\frac{1}{2}\vec{\lambda}^2[\bar{u}_2^\xi(q)-(\bar{u}^\xi_1(q))^2]}[\bar{u}_1^\xi(q_x)-v_{i}(q_x)]^{t_{xi}}[\bar{u}_1^\xi(q_y)-v_{i}(q_y)]^{t_{yi}},
\end{equation}
where $v^{cc}(\mathbf{q})=v_{\ell_{c}\tilde{\ell}_c}(\mathbf{q})$, $v^{cv}(\mathbf{q})=v_{\ell_{c}\tilde{\ell}_v}(\mathbf{q})$, $v^{vc}(\mathbf{q})=v_{\ell_{v}\tilde{\ell}_c}(\mathbf{q})$, and $v^{vv}(\mathbf{q})=v_{\ell_{v}\tilde{\ell}_v}(\mathbf{q})$. To evaluate it numerically, we expand the exponential with $\theta$-dependent into Taylor series and keep track of convergence. 

\section{MoS$_2$ homobilayer with $\theta\approx0$}

In this section, we calculate the spectrum of the MoS$_2$ homobilayer with $\theta\approx0$ in Fig.\ref{fig:MoireXP}. In contrast to the $\theta\approx60^\circ$ case, the attractive interaction in this stacking is stronger because the inversion symmetry forbid large electric dipole moment in a hybridized exciton. 
\begin{figure}
    \centering
    \includegraphics[width=7in]{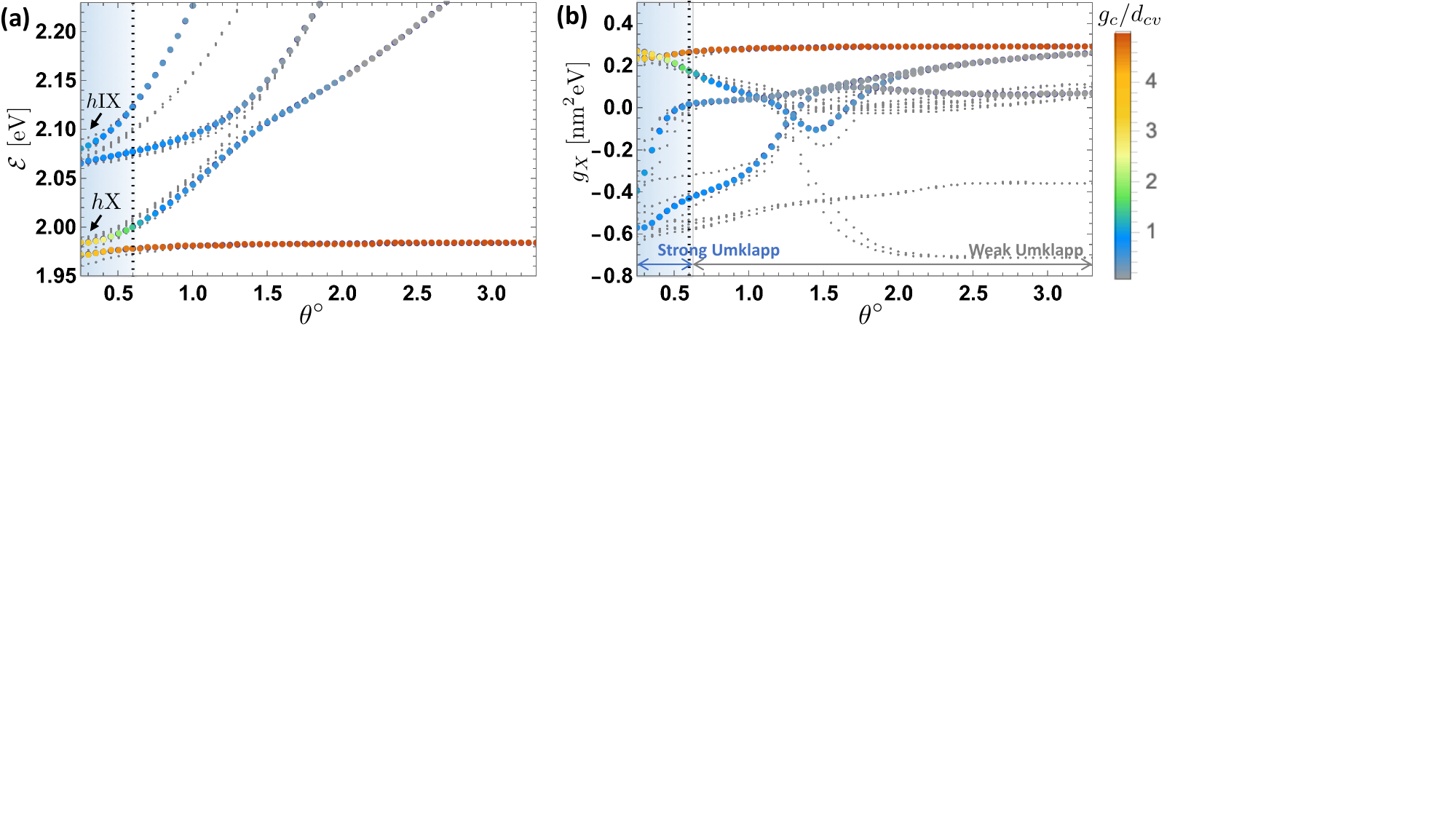}
    \caption{(a) Moir\'e exciton spectrum with $\theta\approx0$. (b) the nonlinearity of moir\'e exciton.}
    \label{fig:MoireXP}
\end{figure}

\section{Calculation of polariton density of state $\rho_c(\omega)$}

To calculate the photonic density of states in the cavity, we employ a mean-field approximation by replacing the interaction term $(\hat{x}^\dagger_{\bar{\alpha}}\hat{x}_{\bar{\alpha}})^2\approx  L^2n_{\bar{\alpha}} \hat{x}^\dagger_{\bar{\alpha}}\hat{x}_{\bar{\alpha}}$ with the pump-power dependent density $n_{\bar{\alpha}}=\langle\hat{x}^\dagger_{\bar{\alpha}}\hat{x}_{\bar{\alpha}}\rangle/L^2$. This yields the mean-field Hamiltonian $\mathcal{H}_{\mathrm{xp}}\approx \Psi^\dagger\bar{H}_{\mathrm{xp}}\Psi$ with the field operator $\Psi^{\dagger}=[c^\dagger_0,\hat{x}^\dagger_{1\sigma},\dots,\hat{x}^\dagger_{\bar{\alpha}\sigma},\dots]$ and the Hamiltonian matrix
\begin{equation}
	\bar{H}_{\mathrm{xp}}=\begin{bmatrix}
	\omega_c-i\kappa&U\\
	U^T&H_X
	\end{bmatrix},\quad 
\end{equation}
where the light-matter coupling vector $[U]_{\bar{\alpha}}=g^{\bar{\alpha}}_c$ and the exciton Hamiltonian matrix is $[H_X]_{\bar{\alpha}\bar{\alpha}'}=(\mathcal{E}^{\bar{\alpha}}_{\sigma}-i\gamma_{\bar{\alpha}}+g^{\bar{\alpha}}_Xn_{\bar{\alpha}})\delta_{\bar{\alpha}\bar{\alpha}'}$. The lifetime of the cavity photon and hybridized moir\'e exciton are $\kappa$ and $\gamma_{\bar{\alpha}}$. 
However, we note that this approximation may not be valid for very large $n_{\bar{\alpha}}$ where a strongly-correlated excitonic states emerge. Thus, we can straightforwardly calculate Green's function $G(\omega)=[\omega-\bar{H}_{\mathrm{xp}}]^{-1}$. This yields $\rho_c(\omega)=\text{Im}[G(\omega)]_{11}$ where the $11$-component corresponds to the photonic space.

\section{Nonlinear absorption spectrum}
The absorption for a exciton with energy $\mathcal{E}$ can be obtained by using the Fermi golden rule with a Lorentz broadening ($\beta=5$~meV), leading to
\begin{equation}
	\Gamma(\mathcal{E})=\frac{2\pi}{\hbar}\sum_{\bar{\alpha}}\frac{(\beta/\pi)|\langle 0|\mathcal{X}^{\bar{\alpha}}_{\sigma}\mathcal{H}_{\mathrm{sc}}c^\dagger_{\mathbf{Q}}| 0\rangle|^2}{[\mathcal{E}^{\bar{\alpha}}_{\sigma}+n_{\bar{\alpha}}g^{\bar{\alpha}}_X-\mathcal{E}(\mathbf{Q})]^2+\beta^2},
\end{equation}
where the photon dispersion is $\mathcal{E}(\mathbf{Q})=\hbar|\mathbf{Q}|$. Also,  $n_{\bar{\alpha}}=\langle\hat{x}^\dagger_{\bar{\alpha}\sigma}\hat{x}_{\bar{\alpha}\sigma}\rangle/L^2$is the excitons density in the background which depends on pump power. This result is used in the plot of Fig.3.

\end{widetext}

\end{document}